\documentclass[twocolumn,showpacs,preprintnumbers,amsmath,amssymb]{revtex4}
\usepackage{graphicx}% Include figure files
\usepackage{dcolumn}% Align table columns on decimal point
\usepackage{bm}% bold math
\usepackage{epsfig}
\def\AuAu{${\rm Au}+{\rm Au}$}
\def\nch{${\rm N^{ch}_{\mid \eta \mid \le 5.4} = 82 \pm 6(syst) }$}
\def\ncht{${\rm N^{ch}_{tot} = 99 }$}
\def\snn{${ \sqrt{s_{_{NN}}}~ = \rm {200~GeV} }$}
\def\sn{$\sqrt{s_{_{NN}}}$}

\def\dAu{${\rm d}+{\rm Au}$}

\def\ppbar{${\rm p}+\bar{\rm p}$}
\def\avgNp{\langle N_{part} \rangle}

\def\avgdn{${\rm \langle dN_{ch}/d\eta \rangle_{\mid \eta \mid \le
0.6} = 9.4 \pm  0.7(syst)}$}

\def\dnch{${\rm dN_{ch}/d{\rm \eta} }$}

\def\avgNp{\langle N_{part} \rangle}
\def\Npp{\langle N_{part}/2 \rangle}

\begin{document}
\title{Pseudorapidity Distribution of Charged Particles 
in \dAu\ Collisions at~\snn }  
\author{
B.B.Back$^1$,
M.D.Baker$^2$,
M.Ballintijn$^4$,
D.S.Barton$^2$,
B.Becker$^2$,
R.R.Betts$^6$,
A.A.Bickley$^7$,
R.Bindel$^7$,
W.Busza$^4$,
A.Carroll$^2$,
M.P.Decowski$^4$,
E.Garc\'{\i}a$^6$,
T.Gburek$^3$,
N.George$^2$,
K.Gulbrandsen$^4$,
S.Gushue$^2$,
C.Halliwell$^6$,
J.Hamblen$^8$,
A.S.Harrington$^8$,
C.Henderson$^4$,
D.J.Hofman$^6$,
R.S.Hollis$^6$,
R.Ho\l y\'{n}ski$^3$,
B.Holzman$^2$,
A.Iordanova$^6$,
E.Johnson$^8$,
J.L.Kane$^4$,
N.Khan$^8$,
P.Kulinich$^4$,
C.M.Kuo$^5$,
J.W.Lee$^4$,
W.T.Lin$^5$,
S.Manly$^8$,
A.C.Mignerey$^7$,
R.Nouicer$^{2,6}$,
A.Olszewski$^3$,
R.Pak$^2$,
I.C.Park$^8$,
H.Pernegger$^4$,
C.Reed$^4$,
C.Roland$^4$,
G.Roland$^4$,
J.Sagerer$^6$,
P.Sarin$^4$,
I.Sedykh$^2$,
W.Skulski$^8$,
C.E.Smith$^6$,
P.Steinberg$^2$,
G.S.F.Stephans$^4$,
A.Sukhanov$^2$,
M.B.Tonjes$^7$,
A.Trzupek$^3$,
C.Vale$^4$,
G.J.van~Nieuwenhuizen$^4$,
R.Verdier$^4$,
G.I.Veres$^4$,
F.L.H.Wolfs$^8$,
B.Wosiek$^3$,
K.Wo\'{z}niak$^3$,
B.Wys\l ouch$^4$,
J.Zhang$^4$\\
\vspace{3mm}
\small
$^1$~Argonne National Laboratory, Argonne, IL 60439-4843, USA\\
$^2$~Brookhaven National Laboratory, Upton, NY 11973-5000, USA\\
$^3$~Institute of Nuclear Physics, Krak\'{o}w, Poland\\
$^4$~Massachusetts Institute of Technology, Cambridge, MA 02139-4307, USA\\
$^5$~National Central University, Chung-Li, Taiwan\\
$^6$~University of Illinois at Chicago, Chicago, IL 60607-7059, USA\\
$^7$~University of Maryland, College Park, MD 20742, USA\\
$^8$~University of Rochester, Rochester, NY 14627, USA\\
}
\date{\today} 
\begin{abstract}
The measured pseudorapidity distribution of primary charged
particles in minimum-bias~\dAu\ collisions at~\snn~is presented for 
the first time. This distribution falls off less rapidly in the
gold direction as compared to the deuteron direction.
The average value of the charged particle pseudorapidity
density at midrapidity 
is~\avgdn~and the integrated primary charged
particle multiplicity in the measured region is 82 $\pm$ 6(syst). Estimates of the total
charged particle production, based on extrapolations outside the
measured pseudorapidity region, are also presented.
The pseudorapidity distribution, normalized to the number of
participants in~\dAu~collisions, is compared to those 
of~\AuAu~and~\ppbar~systems at the same energy.
The~\dAu~ distribution is also compared to the
predictions of the parton saturation model, as well as microscopic models.
\end{abstract}
\maketitle
The pseudorapidity distribution of charged particles
in~\dAu~collisions is
important for understanding the evolution of the system
created in more complicated~\AuAu~collisions, and may provide a constraint on the
initial state parton density. 
The energy and centrality dependence of pseudorapidity distributions
in~\AuAu~collisions measured at RHIC~\cite{BacPRL2001,Bac2,Brah} are 
consistent with the
approach based on the ideas of parton
saturation~\cite{KhPLB523,GriPR100} and semi-classical
QCD~\cite{MacPRD}. Measurements of~\dAu~collisions may be the ideal
way to search for the onset of gluon saturation, since the system
should be
much simpler than the one studied in~\AuAu~collisions. The results
of~\dAu~collisions are therefore crucial for testing the saturation approach \cite{Dima2003}.
\par
In this Letter we present the first measurement of the minimum-bias pseudorapidity 
distribution of primary charged particles~(\dnch)~produced in
collisions of 
deuterons with gold nuclei at a nucleon-nucleon center-of-mass
energy,~\sn, of 200 GeV.  
The pseudorapidity, ${\rm \eta }$, is defined as 
${\rm \eta = -ln[tan(\theta/2)]}$, where ${\rm \theta }$   
is the emission angle relative to the direction of the deuteron beam.
The data were obtained with the 
PHOBOS detector at the Relativistic Heavy Ion Collider at Brookhaven National Laboratory.
The data were collected using the multiplicity
array~\cite{BacNim}, covering ${\rm \mid\eta\mid\le 5.4}$. 
The array consisted of a barrel of silicon detectors surrounding the beam pipe 
in the central rapidity region (``Octagon''), and six forward silicon counters,
three on each side of the interaction point
(``Rings'').  
The multiplicity array used in~\dAu~collisions was the same as that 
for~\AuAu~collisions at~\snn~\cite{BacPRL2003}. The detector setup also included two
sets of 16 scintillator counters, covering ${\rm 3 < \mid \eta \mid < 4.5}$. 
These counters were used in the primary event trigger and in the offline event selection.
\begin{figure}[hbt]
\hspace*{-0.5cm}\epsfig{file=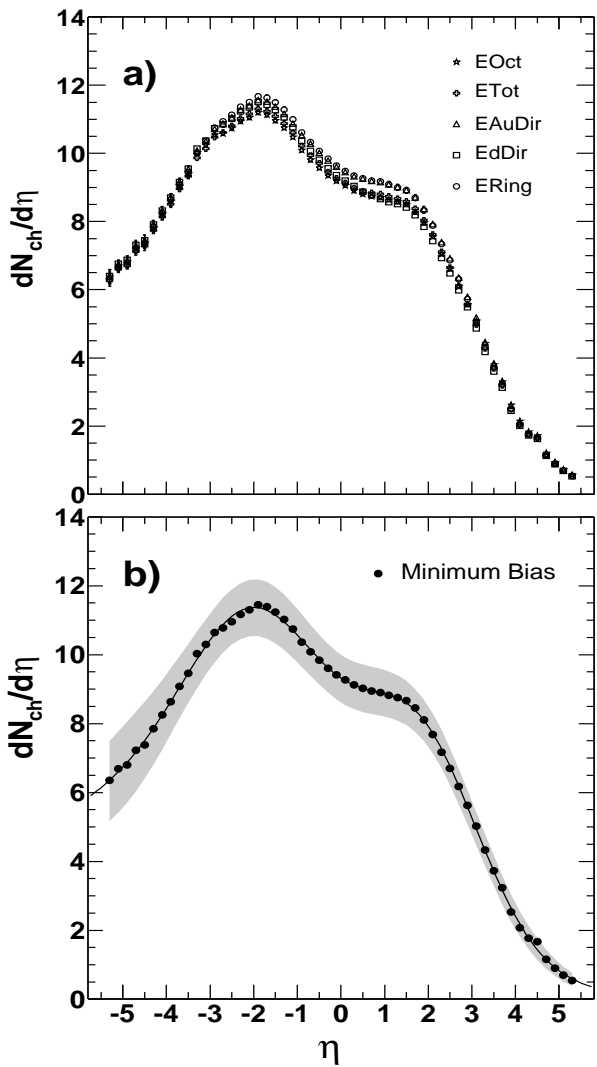,height=15.1cm,width=8cm}
\vspace*{-1.2cm}
\caption{Minimum-bias pseudorapidity distributions of primary charged
particles,~\dnch, measured for~${\rm d + Au}$ at~\snn. a)~\dnch~distributions
obtained by explicit integration over centrality bins for each centrality
measure. b) Final~\dnch~distribution obtained by
averaging over the five minimum-bias distributions. 
The gray band corresponds to the 
systematic errors (90\% C.L.). The curve
corresponds to a triple Gaussian fit to~\dAu.
}
\label{fig:fig1}
\end{figure}

\par The results presented in this Letter were obtained by two analysis
methods: a ``hit-counting'' method and an ``analog'' method.  The
details of the analysis procedure of the two methods leading to the
measurement of~\dnch~can be found in Ref.~\cite{BacPRL2001}.  The
measured pseudorapidity distribution was corrected for particles which
were absorbed or produced in the surrounding material and for feed-down products from
weak decays of neutral strange particles. Uncertainties associated with these
corrections, which we take as 20--50\% of the corrections, dominate
the systematic errors.  
\par Due to the low multiplicity in~\dAu, a
new algorithm for collision vertex reconstruction was developed, using
the hit position and energy deposited in the Octagon
detector. Monte-Carlo studies using HIJING~\cite{HIJING} and GEANT
3.21, as well as comparisons with
the vertex information from reconstructed tracks, show a resolution of
${\rm \sigma_{vtx} (z) = 1.4~cm}$ for the most peripheral events and
${\rm \sigma_{vtx} (z) = 0.8~cm}$ for the most central events, 
$z$ being the distance along the beam axis. 

In order to determine the minimum-bias pseudorapidity distribution for
this analysis, it is necessary to correct for the trigger and vertex
finding efficiencies. Since these efficiencies are functions of
multiplicity (and therefore centrality), the most model-independent
approach is to measure the pseudorapidity distribution in narrow bins
of centrality and then integrate over centrality to produce a minimum-bias
result. The alternative approach of multiplying by a global Monte-Carlo
(MC) correction factor (reconstructed/true MC) would rely on our
model to get the details of the shape and the centrality dependence right.
To further reduce the model-dependence, we used five distinct measures for
estimating the collision centrality based on the multiplicity (defined by
the energy deposited in the silicon detectors) in different regions of
pseudorapidity. The first of these centrality measures, ${\rm
E_{Tot}}$, used data from the full pseudorapidity coverage of PHOBOS,
${ \rm \mid \eta \mid \le 5.4}$.  The second measure, ${\rm
E_{Oct}}$, used data from the more central region ${\rm \mid\eta\mid
\le 3}$. The next two measures, ${\rm E_{AuDir}}$ and ${\rm E_{dDir}}$, used
data from the gold direction (${ \rm -5.4 \le \eta \le -0.5}$) and
deuteron direction (${ \rm 0.5 \le \eta \le 5.4 }$),
respectively. The final centrality measure, ${\rm E_{Ring}}$, used data
far from the mid-rapidity region from the silicon~ring~counters at~${3.0
\le \rm \mid \eta \mid \le 5.4}$.  
\par 
The centrality measures were calibrated using HIJING~\cite{HIJING} and GEANT
3.21 simulations. In the case of ${\rm E_{Tot}}$, ${\rm E_{Oct}}$, ${\rm E_{AuDir}}$ and
${\rm E_{dDir}}$, the distributions in data and MC were very similar once the
triggering and event selection were applied to the MC and an additional
scaling factor near unity was included. For instance, for ${\rm E_{Tot}}$ this
scaling factor was 1.046. This allowed us to use the zero-bias ``true''
distribution in the MC to divide the data into equal 10\% cross-section
bins and to estimate the overall trigger and vertex-finding efficiency of
 $\sim$ 83\%. Based on studies of reconstructed and true pseudorapidity
distributions from different Monte-Carlo simulations, we include a global 5$\%$ 
systematic uncertainty on the pseudorapidity distribution in the reported errors.  
In the case of ${\rm E_{Ring}}$, the MC and data distributions were
sufficiently different that the procedure had to be modified. 
From the calculated efficiencies in each of the MC centrality bins, we
could derive how many of the events in each ``true'' 10\% bin would
appear in the data. Cuts were then made on the ${\rm E_{Ring}}$
distribution that gave this efficiency-corrected number of events. 
\par
Once the efficiency had been calculated and centrality cuts made, we
extracted the average number of participating nucleons $\avgNp$ for
each centrality bin using HIJING+GEANT. 
The measured minimum-bias pseudorapidity distribution and the estimated
number of participants were obtained by averaging
over all centralities for each centrality measure. This averaging
was done as the minimum-bias distribution is less sensitive to
centrality resolution and multiplicity bias effects than the
individual centrality bins. The final distribution was obtained by
averaging over the five minimum-bias distributions given by the
different centrality measures.
The total number of participant nucleons,   
$\avgNp$, was estimated to be ${\rm 8.1 \pm 0.7(syst) }$.
To cross check the final distribution
obtained by using the silicon centrality measures, a further analysis was
performed, only requiring at least one hit in one of the scintillator
counter arrays. This analysis was consistent within the present
systematic uncertainty.

\begin{figure}[htbp]
\hspace*{-0.5cm}\epsfig{file=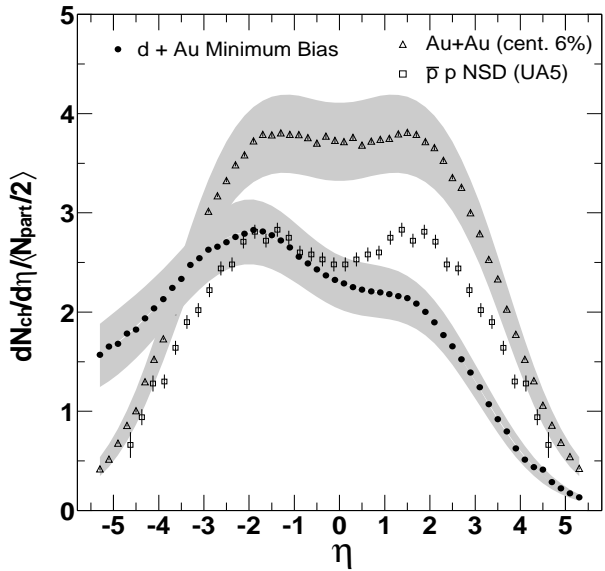,height=8.cm,width=8.cm}
\vspace*{-0.4cm}
\caption{$dN_{ch}/d\eta$ distributions per participant are shown for 
minimum-bias~\dAu~collisions at~\snn (solid points), for central~\AuAu~collisions
(0--6$\%$)~\cite{BacPRL2003} (triangles) and for~\ppbar~collisions
from UA5~\cite{UA5} (open squares) at the same energy. 
The systematic errors are shown by gray bands. 
The uncertainties on $\avgNp$ for~\dAu~and~\AuAu~
have been added in quadrature to the gray bands.
}
\label{fig:fig2}
\end{figure}

\begin{figure}[htbp] 
\hspace*{-0.5cm}\epsfig{file=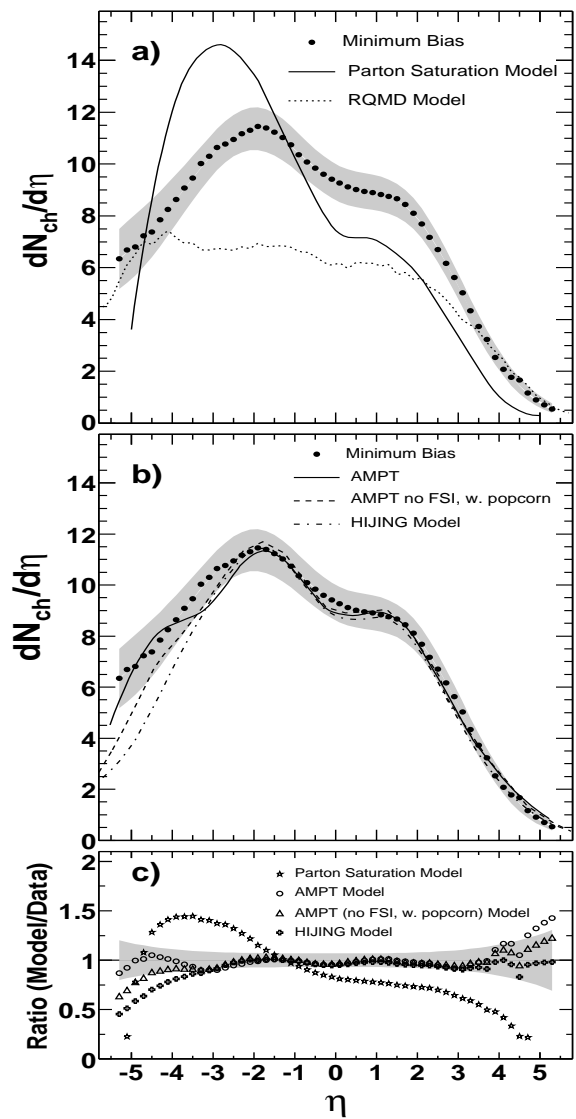,height=15.0cm,width=7.5cm}
\vspace*{-0.1cm}
\caption{Comparison of the measured minimum-bias pseudorapidity 
distribution for~\dAu~collisions at \snn~to model predictions. a)
Comparison to the parton saturation~\cite{Dima2003} and RQMD~\cite{RQMD} models. 
b)~Comparison to the
predictions of the HIJING~\cite{HIJING} model and the published AMPT model with and without final-state
interactions (FSI)~\cite{AMPT}. c)~Quantitative evaluation
of the model predictions, expressed as
the ratio of the model prediction to the data.  The gray band corresponds
to the systematic errors (90\% C.L.) on the data. 
}
\label{fig:fig3}
\end{figure}
\par 
Figure~\ref{fig:fig1}a shows the minimum-bias pseudorapidity
distributions of primary charged particles measured for~${\rm d + Au}$
collisions at~\snn~obtained with the five
centrality measures described above. The distributions agree
to within~${\rm 5\%}$ everywhere. Fig.~\ref{fig:fig1}b shows the
final result obtained by averaging with equal weights over the five
distributions in Fig.~\ref{fig:fig1}a. The systematic errors (90$\%$
C.L.) are shown as a gray band. The statistical errors are 
negligible. Fig.~\ref{fig:fig1}b clearly shows
that the cross section for particle production is largest in
the gold direction (${\rm \eta < 0}$), and smallest in
the deuteron direction (${\rm \eta > 0}$). 
 By fitting the gold-side
using a Gaussian in the region ${\rm -3 \le \eta \le -0.5 }$, the peak
centroid was found to be at
${\rm \eta \sim -1.9}$. This peak position is much closer to
mid-rapidity than predicted by the parton saturation 
model \cite{Dima2003}, see Fig.~3a. It may be interesting to note that
a double Gaussian decomposition 
of the Au+Au distribution at~\snn~\cite{BacPRL2003} for the ${\rm 0-6\%}$ most central
collisions places the Gaussian centroids at ${\rm \eta \sim \pm1.9}$,
see Fig.~2. 
The significance of this observation is, however, not obvious, and may indeed be accidental.
\par
For~\dAu, the measured average value of the charged particle pseudorapidity
density at midrapidity 
is~\avgdn~and the integrated primary charged
particle multiplicity in the measured region is~\nch. 
Since ${\rm dN_{ch}/d\eta}$ falls off to only about half of its maximum
value at the most negative pseudorapidity of the detector acceptance,
it is non-trivial to estimate the total charged particle
multiplicity, ${\rm N_{ch}}$, for this system.  However, using a triple
Gaussian fit to the data we
obtain the solid curve in Fig.~\ref{fig:fig1}b, which corresponds 
to a total number of produced charged particles of~\ncht~and the fit of
the upper limit of the systematic errors (upper limit of the gray
band) gives ${\rm N_{ch} = 110}$. 
We believe, however, that this represents an extreme upper limit for
${\rm N_{ch}}$ since experimental data on p+A multiplicity distributions at
lower energies \cite{wit1,wit2,wit3,wit4,wit5}, as well as various theoretical
models discussed below, indicate that the fall-off at large negative
pseudorapidities is more abrupt than indicated by this fit to the
data. A more reasonable extrapolation, guided by the expectations
based on lower energy p+A data and predictions of the AMPT model,
yields an estimate of ${\rm N_{ch}= 5}$ in the extrapolated region. 
Therefore, we estimate that the total charged particle
multiplicity 
in this reaction is ${\rm N_{ch}=87^{+23}_{-07}}$. It should be kept in mind,
however, 
that even this estimate contains a significant contribution from the unmeasured region of pseudorapidity.     
\par
Figure~\ref{fig:fig2} shows the normalized minimum-bias \dnch~
distribution for~\dAu~in comparison to the most
central (0--6\%)~\AuAu~collisions~\cite{BacPRL2003}   
and non-single-diffractive~\ppbar~collisions from UA5~\cite{UA5} 
at the same energy,~\snn. To be consistent with our previous Au + Au 
pseudorapidity distribution publications, we have used a normalization factor
of $\Npp$ which is unity for~\ppbar. The distributions from different
systems are normalized chiefly to enable a comparison of their shapes. 
The distribution seen in d+Au collisions at RHIC has features similar to those seen in
p+A collisions at lower
energies~\cite{wit1,wit2,wit3,wit4,wit5}. Compared
to~\ppbar~collisions, 
there is a significant increase in particle
production in the gold fragmentation region and a reduction of
particle production at positive pseudorapidity in the deuteron
direction. The overall normalized 
production of particles is approximately the same
as in proton-(anti-) proton collisions~\cite{UA5}.

\par
Figure~\ref{fig:fig3}a shows a comparison of the measured minimum-bias
pseudorapidity distribution to the published predictions of the parton saturation
model~\cite{Dima2003} and the results from RQMD~\cite{RQMD}.
It is evident that 
these models are inconsistent with the data. The parton saturation model
overestimates the height of the gold-side peak, underestimates its
width, and predicts the peak at ${\rm \eta \sim -3}$ rather than ${\rm
\eta = -1.9}$ as in the data. On the deuteron side, the model 
underestimates the charged particle production.
RQMD drastically underpredicts the particle production and has
surprisingly little pseudorapidity asymmetry.
\par 
Figure~\ref{fig:fig3}b presents the
comparison of the measured~\dnch~distribution to the predictions 
of HIJING~\cite{HIJING} and the published AMPT
calculation~\cite{AMPT}. 
The HIJING calculation (dash-dotted curve) reproduces the deuteron side and the peak of the gold-side,
but fails to reproduce the tail in the gold direction (${\rm \eta < -2.5}$). 
The AMPT model uses HIJING for the initial stage of the collision,
adding  a ``popcorn mechanism'' for
baryon-antibaryon production~\cite{AMPT,AMPT1} and final-state interactions (FSI). 
The solid curve is
the default, while the dashed curve excludes FSI. 
We see that both final-state interactions and
the popcorn mechanism appear to broaden the gold-side peak,
leading to a moderate increase of the particle
multiplicity in the region ${\rm \eta \le -3.5}$. 
The ratio of the model predictions to data is shown in
Fig.~\ref{fig:fig3}c.

\par
In summary, the pseudorapidity distribution of charged particles
produced in~\dAu~collisions at~\snn~ has been measured.
The distribution is rather broad and peaked in the gold direction 
similar to observations in p+A collisions at lower collision energies.
The measured pseudorapidity distribution is compared with the
predictions of the parton saturation model as well as microscopic models.
While AMPT predictions fall close to the data, the saturation model overestimates 
the asymmetry between particle production in the gold and deuteron hemispheres. 
\newline
\par
We thank RHIC Operations for providing the variety of colliding
systems. This work was partially supported by U.S. DOE grants DE-AC02-98CH10886,
DE-FG02-93ER40802, DE-FC02-94ER40818, DE-FG02-94ER40865, DE-FG02-99ER41099,
and W-31-109-ENG-38, US NSF grants 9603486, 9722606 and 0072204, Polish KBN
grant 2-P03B-10323, and NSC of Taiwan contract NSC 89-2112-M-008-024.

\end{document}